\documentclass[twoside,hidelinks,11pt]{article}

\usepackage{jmlr2e} 

\usepackage[dvipsnames]{xcolor}
\usepackage{amsmath}
\usepackage{tikz}
\usepackage{xfrac}
\usepackage{float}
\usepackage{hyperref}

\usepackage{GLAGraphs}

\newcommand{\mat}[1]{\mathbf{#1}}

\definecolor{cumulantscolor}{RGB}{84,39,143}
\definecolor{momentscolor}{RGB}{0,109,44}
\definecolor{unknowncolor}{RGB}{230,85,13}

\newcommand{\shrinkA}{\hspace{-1.75pt}}
\newcommand{\shrinkB}{\hspace{-1pt}}

\newcommand{\deltapi}{\delta\kern-1pt\boldsymbol{\pi}}
\newcommand{\deltab}{\delta\kern-1pt\mat{B}}
\newcommand{\pae}{\mathrm{P\!A\!E}}
\newcommand{\pare}{\mathrm{P\!A\!R\!E}}

\newcommand{\ER}{\mathrm{E\kern-1pt R}}
\newcommand{\fallingfact}[2]{{#1}^{\underline{#2}}_{\vphantom{1}}}
\newcommand{\ExpLeftPar}{\scalebox{0.55}{(}}
\newcommand{\ExpRightPar}{\scalebox{0.55}{)}}

\newcommand{\equalAuthorString}{\textit{Shared first authorship.}}

\newcommand{\urlcolor}{blue}
\newcommand{\internallinkcolor}{black} 
\newcommand{\citationcolor}{blue}
\hypersetup{
  colorlinks   = true, 
  urlcolor     = \urlcolor, 
  linkcolor    = \internallinkcolor, 
  citecolor   = \citationcolor 
}
\newcommand{\gundyast}{Gunderson$^\ast$}
\newcommand{\bhast}{\mbox{Bravo-Hermsdorff}$^\ast$}

\newcommand{\TitleString}{Quantifying Network Similarity using Graph Cumulants}

\newcommand{\ShortTitleString}{Quantifying Network Similarity using Graph Cumulants}

\usepackage{lastpage}
\jmlrheading{24}{2023}{1-\pageref{LastPage}}{7/21; Revised
1/23}{6/23}{21-0828}{Gecia \bhast, Lee M.~\gundyast, \mbox{Pierre-Andr\'e} Maugis, and Carey E.~Priebe}

\ShortHeadings{\ShortTitleString}{\bhast, \gundyast, Maugis, and Priebe}
\firstpageno{1}

\begin{document}
\title{\TitleString}
    \author{\name Gecia \mbox{Bravo-Hermsdorff}\kern1.5pt\thanks{ \equalAuthorString} \email gecia.bravo@gmail.com \\
             \addr Department of Statistical Sciences\\
             University College London\\
    London, WC1E 7H8, United Kingdom
          \AND
                   \name Lee M.~Gunderson\kern0.0pt\gundyast \email l.gunderson@ucl.ac.uk\\
          \addr Gatsby Computational Neuroscience Unit\\
         	University College London\\
         London, W1T 4JG, United Kingdom
    \AND
    \name \mbox{Pierre-Andr\'e} Maugis \email pamaugis@google.com \\
          \addr Google Research \\
          Z{\"u}rich, 8002, Switzerland
    \AND 
    \name Carey E.~Priebe \email cep@jhu.edu \\
          \addr Department of Applied Mathematics and Statistics \\
                  Whiting School of Engineering \\
                  Johns Hopkins University \\
                  Baltimore, MD 21218, USA
    }
\editor{Tina Eliassi-Rad}

\maketitle

\begin{abstract}
How might one test the hypothesis that networks were sampled from the same distribution?
Here, we compare two statistical tests that use subgraph counts to address this question.
The first uses the empirical subgraph densities themselves as estimates of those of the underlying distribution.  
The second test uses a new approach that converts these subgraph densities into estimates of the \textit{graph cumulants} of the distribution 
(without any increase in computational complexity).  
We demonstrate --- via theory, simulation, and application to real data --- the superior statistical power of using graph cumulants. 
In summary, when analyzing data using subgraph/motif densities, we suggest using the corresponding graph cumulants instead. 
\end{abstract}


\section{Statement of the Problem}
\label{Sec:ProblemStatement}
In this paper, we use statistics based on subgraph counts to address the following problem:

\begin{center}
\fbox{
\parbox{0.74\textwidth}{
Given two samples, \mbox{$\boldsymbol{G}_{\shrinkA A}^{ }$} and \mbox{$\boldsymbol{G}_{\shrinkB B}^{ }$}, each containing $s$ graphs\\
sampled i.i.d.~from unknown distributions \mbox{$\mathcal{G}_{\shrinkA A}^{ }$} and \mbox{$\mathcal{G}_{\shrinkB B}^{ }$}, respectively,\\
the goal is to infer whether \mbox{$\mathcal{G}_{\shrinkA A}^{ }$} and \mbox{$\mathcal{G}_{\shrinkB B}^{ }$} are different distributions. 
}}
\end{center}

\newpage

\section{What we are Doing (Motivation)}
\label{Sec:AreDoing}
The theory for statistical analysis of i.i.d.~data (e.g., the height and weight of different breeds of dogs) is \mbox{well-established} \citep{casella2021statistical,stuart1963advanced}.  
However, when the data of interest are \textit{interactions} (e.g., who plays with whom in the dog park), a similar consensus has not yet been reached \citep{orbanz2017subsampling}.  

To analyze such networks of interactions, one common approach is to use statistics based on counts of small substructures (i.e., subgraphs or ``motifs'')  \citep{aliakbarpour2018sublinear,maugis2020testing,pattillo2013clique,xia2019survey}.  
For example: the densities of \mbox{``k-star''} subgraphs provide increasingly detailed information about a network's degree distribution \citep{rauh2017polytope}, and the density of cliques of various sizes gives information about the scale of its clustering \citep{ouyang2019clique}.  

To make meaningful comparisons, one must first answer the question: ``Compared to what?'' 
The \mbox{frequently-used} clustering coefficient opts to divide the number of ``complete'' triangles by the number of ``incomplete'' triangles \citep{newman2001structure}.  
The \mbox{often-cited} configuration model compares the observed subgraph counts with those of random networks with the same degree distribution \citep{alon2007network, barabasi1999emergence}. 

The \mbox{recently-proposed} graph cumulants \citep{Gunderson2019Introducing} offer a natural set of statistics based on a combinatorial view of cumulants (e.g., mean, variance, skew, kurtosis, etc.).  
Graph cumulants compare the density of a substructure in a network to the density that would be expected due to the prevalence of smaller substructures, thereby quantifying an intuitive notion of ``excess propensity'' for that substructure.

It is \mbox{well-known} that subgraph densities are useful for characterizing individual networks \citep{bickel2011method, bhattacharyya2015subsampling} and for comparing different networks \citep{maugis2020testing, bhattacharya2022motif, zhang2022edgeworth, jin2021optimal}.  
A few works have also used statistics that combine subgraph densities to compare networks.  
In particular, \citet{gao2017testing} uses two statistics (one involving edge and \mbox{\(2\)-star} densities, and the other involving edge and triangle densities) to devise a test for distinguishing an \mbox{Erd\H{o}s-R\'enyi} model from an alternative stochastic block model (SBM).

Here, we describe a simple statistical test for comparing networks based on graph cumulants.  
For comparison, we consider the analogous test based on graph moments (i.e., subgraph densities) \citep{maugis2020testing}.  
Our results strongly suggest that graph cumulants should be the default choice of statistics for problems involving subgraph counts.

\section{What we are Not Doing (Related Work)}
\label{Sec:NotDoing}
There are many ways to compare networks \citep{tantardini2019comparing}.  
Here, we describe some other common approaches,
highlighting their differences with the setting considered in this paper.

\subsection{Spectral Methods}
Broadly, these methods use the eigendecomposition of matrices associated with the network, such as the adjacency matrix or the graph Laplacian \citep{chen2020spectral, grav2019aunifying, mukherjee2016clustering}.  
Here, we focus instead on statistics based on the frequency of small subgraphs. 
These two approaches are complementary \citep{lovasz2012large}; 
flavorfully, spectral methods are more sensitive to the graph's global structure (its ``shape''), whereas methods based on subgraph frequencies are more sensitive to the graph's local structure (its ``texture'').

\subsection{Matching Nodes} 
When the networks being compared have the same set of unique names for all their nodes \citep{ghoshdastidar2020two,tang2017semiparametric}, it is highly advantageous for statistical tests to incorporate this \mbox{one-to-one} mapping (e.g., when comparing fMRI data of different subjects, it helps to assume that their hippocampi are sufficiently analogous).  
However, such information is not always available (e.g., when comparing the social interactions within different schools).  
This work addresses the latter problem; the nodes are considered to be indistinguishable (i.e., exchangeable), and only the statistics of their pairwise interactions (i.e., their edges) are known to the tests.

\subsection{Obtaining Significance by Sampling} 
Many tests for comparing networks require sampling from the inferred distributions to estimate the significance of their differences.  
Common examples include the use of configuration models \citep{masuda2018configuration}, exponential random graph models \citep{an2016fitting}, and geometric random graph models \citep{asta2014geometric}.  
Such methods are typically computationally intensive, rendering them difficult (or impossible) to implement in practice \citep{ginoza2010network}.  
The two tests considered in this paper sidestep this issue entirely, analytically computing the significance of the observed differences between the networks.

\section{How we Name Things (Notation)}
\label{Sec:Notation}
A capital $G$ denotes a single graph with $n$ nodes.  
All graphs are assumed to be undirected, unweighted, simple graphs.\footnote{This is for simplicity of presentation; the tests we describe here naturally extend to networks with additional information, such as directed edges, weighted edges, and node attributes  \citep{Gunderson2019Introducing}.}  
A calligraphic $\mathcal{G}$ denotes a distribution over such graphs.  
All graph distributions are assumed to be generated by a single graphon \citep{lovasz2012large} (see Section~\ref{Sec:StatTestMeanAndCov}).\footnote{Again, this assumption may be relaxed.}  
A bold $\boldsymbol{G}$ denotes a sample of $s$ graphs from such a distribution. 
All graphs in a sample are assumed to have the same number of nodes.  

A plebeian $g$ denotes a subgraph.  
An emboldened $\boldsymbol{g}$ denotes a set of subgraphs. 
The set of subgraphs with at most $r$ edges is denoted by \mbox{$\boldsymbol{g}_r^{ }$}, and  
the restriction to connected subgraphs is denoted by \mbox{\smash{$\boldsymbol{g}_{r}^{\ExpLeftPar\textrm{c}\ExpRightPar}$}} (e.g., \mbox{\smash{$\boldsymbol{g}_3^{ } = \{\EdgeBig, \WedgeBig, \EdgeEdgeBig , \TriangleBig, \ClawBig, \ThreelineBig, \EdgeWedgeBig, \EdgeEdgeEdgeBig\}$}} and \mbox{\smash{$\boldsymbol{g}_{3}^{\ExpLeftPar\textrm{c}\ExpRightPar} = \{\EdgeBig, \WedgeBig, \TriangleBig, \ClawBig, \ThreelineBig \}$}}). 
Parenthetical superscripts are also used to distinguish the covariance
matrices of graph moments $\boldsymbol{\Sigma}^{\ExpLeftPar\mu\ExpRightPar}_{\vphantom{1}}$ from those of graph cumulants $\boldsymbol{\Sigma}^{\ExpLeftPar\kappa\ExpRightPar}_{\vphantom{1}}$.

The statistics we consider here, namely, graph moments and graph cumulants, have associated with them a particular subgraph $g$ (see Section~\ref{Sec:StatisticsWeCompare}). 
\mbox{$\mu_g^{ }(\mathcal{G})$} denotes the graph moment (associated with subgraph $g$) of a distribution $\mathcal{G}$ (see Section~\ref{Sec:GraphMoments}), and \mbox{$\kappa_g^{ }(\mathcal{G})$} the corresponding graph cumulant (see Section~\ref{Sec:GraphCumulants}).  
A bold $\boldsymbol{\mu}$ (or $\boldsymbol{\kappa}$) denotes a vector of graph moments (or cumulants), one for each subgraph in $\boldsymbol{g}$.  

The sample estimators of these quantities are given a hat (e.g., $\widehat{\boldsymbol{\mu}}(\boldsymbol{G})$ and $\widehat{\boldsymbol{\kappa}}(\boldsymbol{G})$, see Section~\ref{Sec:StatTest}). 
Expectation is denoted by angled brackets (e.g., for the unbiased estimators of cumulants \mbox{$\big\langle \widehat{\boldsymbol{\kappa}}(\boldsymbol{G})\big\rangle_{\boldsymbol{G}\sim\mathcal{G}}^{ } = \boldsymbol{\kappa}(\mathcal{G})$}, see Section~\ref{SubSec:StatTestMean_Cumulants}).

\section{Two Statistics based on Subgraphs (What we Measure)}
\label{Sec:StatisticsWeCompare}
We first define graph moments (Section~\ref{Sec:GraphMoments}), the statistics to which we compare our new method.  
Then we describe how to convert these to graph cumulants (Section~\ref{Sec:GraphCumulants}), the statistics used by our new method.

\subsection{Graph Moments (The \mbox{Typically-used} Statistics)}
\label{Sec:GraphMoments}
When discussing counts of a subgraph $g$ in some larger graph $G$, 
it is important to distinguish between \textit{induced} counts and \textit{homomorphism} counts \citep{chen2008understanding}; here we are using the latter.  
Another important distinction is that we are using \textit{injective} homomorphism counts.  
To obtain these counts, first consider all mappings from the nodes of $g$ to the nodes of $G$.  

\textit{Injective} refers to a condition on these node mappings: consider only those mappings that do not send different nodes in $g$ to the same node in $G$.  

\textit{Homomorphism} refers to how we decide which of those (injective) mappings are ``counts'': those for which $G$ has edges at all the locations where there are edges in $g$ (i.e., it is still a count even if there are additional edges in $G$).

Out of all injective mappings from the nodes of $g$ to the nodes of $G$, 
the fraction that are ``counts'' is known as the injective homomorphism \textit{density} \citep{lovasz2012large}.  
These densities are the graph moments of a graph $G$, denoted by \mbox{$\mu_g^{ }(G)$}.

\subsection{Graph Cumulants (The New Statistics)}
\label{Sec:GraphCumulants}
Graph cumulants were recently introduced by \citet{Gunderson2019Introducing}. 
We first review the defining features of cumulants, highlighting a \mbox{less-well-known} combinatorial definition (Section~\ref{Subsec:ClassicalCumulants}). 
We then describe the analogue for graphs (Section~\ref{Subsec:GraphCumulants}). 

\subsubsection{Combinatorial Cumulants (Background)}
\label{Subsec:ClassicalCumulants}
First, consider a \mbox{scalar-valued} random variable \mbox{$X \in \mathbb{R}$} sampled from some distribution $\mathcal{X}$.  
The \mbox{$r^{\text{th}}$-order} moment of $\mathcal{X}$ is the expectation of the $r^{\text{th}}$ power of $X$: 
\mbox{$\mu_r^{ }(\mathcal{X}) = \langle X^r_{\vphantom{1}} \rangle_{X\sim\mathcal{X}}^{ }$}.  
These moments may be combined into certain polynomial expressions, 
known as the cumulants $\kappa_r^{ }$ 
(e.g., mean, variance, skew, kurtosis, etc.).  

Cumulants have a uniquely defining property related to sums of independent random variables \citep{rota2000combinatorics}: 
the cumulants of the resulting sum are equal to the sum of the cumulants of those independent random variables (e.g., \mbox{$\textrm{Var}(X + Y) = \textrm{Var}(X) + \textrm{Var}(Y)$}, when $X$ and $Y$ are independent).  
This is essentially the reason behind the central limit theorem 
and the ubiquity of the Gaussian distribution \citep{gnedenko1949limit, hald2000early}.

While the classical cumulants are often defined via the cumulant generating function,  
they also have an equivalent combinatorial definition in terms of a M\"obius transform (see Section~\ref{Sec:PostScript}) \citep{kardar2007statistical, mccullagh2018tensor, lehner2004cumulants}. 
Namely, the $r^{\text{th}}$ moment can be expressed as a sum of cumulants of order $r$ and lower, corresponding to all partitions of $r$ unique elements (see Figure~\ref{Fig:SchematicGraphCumulants}, top three rows).  
In particular, for our \mbox{scalar-valued} example: 
\begin{align}
\mu_r^{ }(\mathcal{X}) = \sum_{\pi \in P_r^{ }} \prod_{b \in \pi} \kappa_{|b|}^{ }(\mathcal{X}), \label{eq:CombinatorialMomentsDef}
\end{align}
where \mbox{$\mu_r^{ }$} is the \mbox{$r^{\text{th}}$} moment, \mbox{$\kappa_r^{ }$} is the \mbox{$r^{\text{th}}$} cumulant, \mbox{$P_r^{ }$} is the set of all partitions of $r$ unique elements (i.e., a set of non-overlapping subsets, or ``blocks'', whose union contains all $r$ elements), $\pi$ is one such partition, $b$ is a block in partition $\pi$, and $|b|$ is the number of elements in block $b$.  

Equation~\ref{eq:CombinatorialMomentsDef} may be rearranged to obtain expressions for the cumulants in terms of moments.  
For example, the third (classical) cumulant (the ``skew'') is: \mbox{$\kappa_3^{ } = \mu_3^{ } - 3\mu_2^{ }\mu_1^{ } + 2\mu_1^3$}.  
This combinatorial definition makes the generalization to random variables with additional structure, such as graphs, more transparent. 

\subsubsection{Graph Cumulants (Definition)}
\label{Subsec:GraphCumulants}
Before describing graph cumulants, it is worth mentioning a subtle point.  
Notice that \mbox{Equation~\ref{eq:CombinatorialMomentsDef}} relates the moments and cumulants of the \textit{distribution} $\mathcal{X}$ (and \textit{not} of any finite sample $\mathbf{X}$).  
While this distinction is somewhat pedantic for graph moments (see Section~\ref{SubSec:StatTestMean_Moments}), 
the combinatorial definition of cumulants (classical or graphical) should only be applied to distributions.  

The moments and cumulants of \mbox{real-valued} random variables are indexed by their order \mbox{$r\in\mathbb{N}$}. 
For \mbox{graph-valued} random variables, moments \citep{lovasz2010graph, bickel2011method} and cumulants \citep{Gunderson2019Introducing} are now indexed by subgraphs \mbox{$g\in\boldsymbol{g}$}, with order given by the number of edges in $g$.  

\begin{figure}[H]
\begin{center}
\centerline{\includegraphics[trim={0cm 0cm 0cm 0cm},clip, width=0.9\columnwidth]{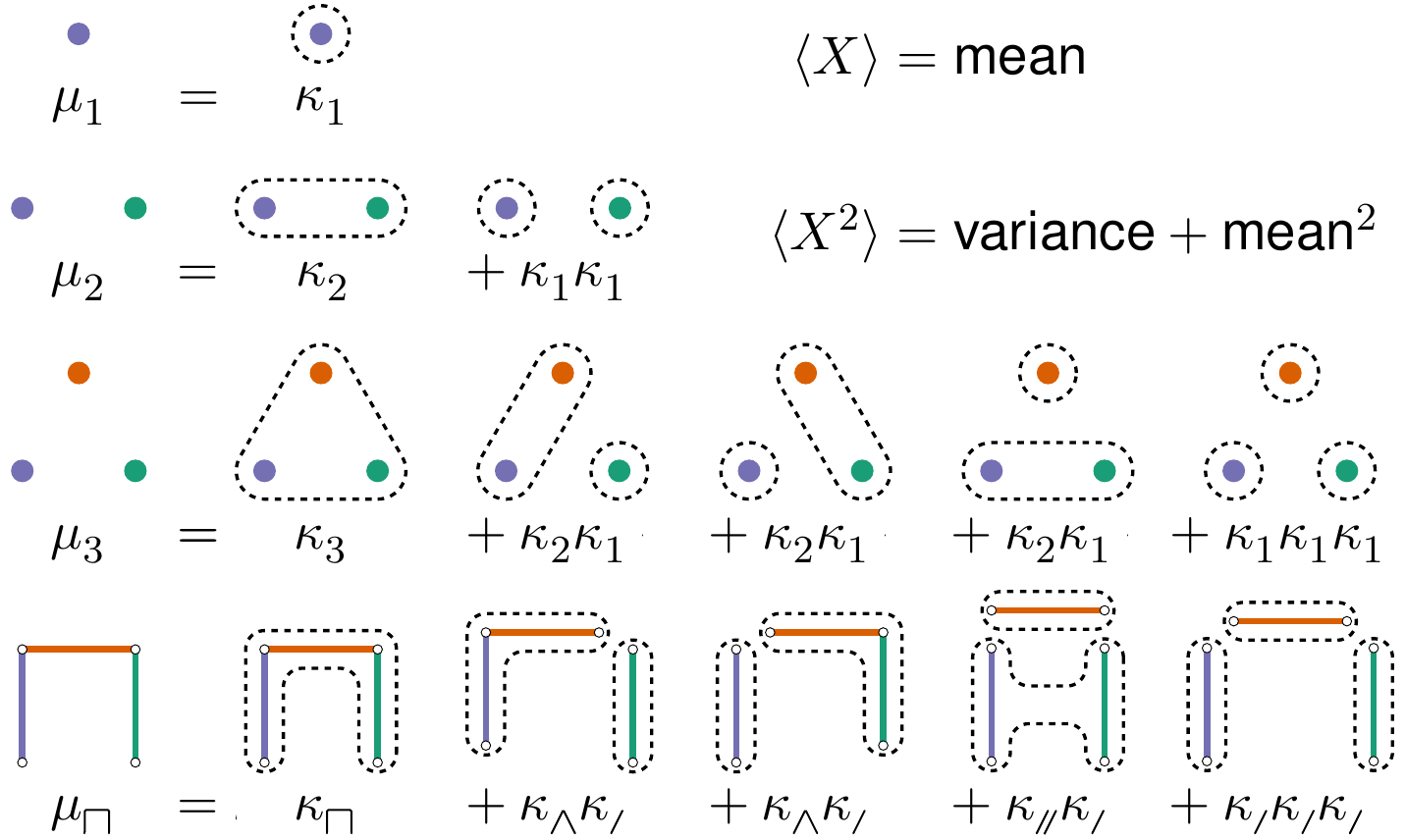}} 
\vspace{-2pt}
\caption{
\mbox{\textbf{To expand a graph moment \mbox{$\mu_{g}^{ }$} in terms of graph cumulants,}} 
\mbox{\textbf{enumerate all partitions of the edges forming subgraph $g$.}} \mbox{\hspace{0.10\textwidth}} 
The top three rows illustrate the combinatorial expansion of the first three classical moments in terms of cumulants (Equation~\ref{eq:CombinatorialMomentsDef}).  
Analogously, the bottom row shows how to expand the graph moment \mbox{$\mu_{\protect\threeline}^{ }$} in terms of graph cumulants (Equation~\ref{eq:CombinatorialGraphMomentsDef}).    
The last term (\mbox{$\kappa_{\protect\oneedge}^{3}$}) corresponds to partitioning this subgraph into three subsets, each with a single edge. 
The first term (\mbox{$\kappa_{\protect\threeline}^{ }$}) corresponds to ``partitioning'' this subgraph into a single subset with all three edges, thus inheriting the connectivity of the entire subgraph.  
The remaining terms (\mbox{$\kappa_{2}^{ }\kappa_{1}^{ }$}) correspond to partitioning this subgraph into a subset with one edge and a subset with two edges.  
This can be done in three different ways: 
in two cases (the two \mbox{$\kappa_{\protect\twowedge}^{ }\kappa_{\protect\oneedge}^{ }$} terms), the subset with two edges has those edges sharing a node; 
and in one case (the \mbox{$\kappa_{\protect\twoparallel}^{ }\kappa_{\protect\oneedge}^{ }$} term), the subset with two edges has those edges not sharing any node. 
}
\label{Fig:SchematicGraphCumulants}
\end{center}
\end{figure}

To apply the combinatorial definition (\mbox{Equation~\ref{eq:CombinatorialMomentsDef}}) to networks, the partitioning \mbox{$P_r^{ }$} of the edges of the subgraphs must respect their connectivity (see Figure~\ref{Fig:SchematicGraphCumulants}, bottom row), i.e.: 
\begin{align}
\mu_{g}^{ }(\mathcal{G}) &= \sum_{\pi \in P_{E(g)}^{ }} \prod_{b \in \pi} \kappa_{g_b^{ }}^{ }(\mathcal{G}), \label{eq:CombinatorialGraphMomentsDef}
\end{align}
where $E(g)$ is the set of edges forming subgraph $g$, \mbox{$P_{E(g)}^{ }$} is the set of all partitions of these edges, and \mbox{$g_b^{ }$} is the subgraph formed by the edges in $b$.  

Again, these may be rearranged to obtain the graph cumulants \mbox{\smash{$\kappa_{g}^{ }(\mathcal{G})$}} in terms of the graph moments \mbox{\smash{$\mu_{g}^{ }(\mathcal{G})$}}. 
For example, the (\mbox{$3^{\text{rd}}$-order}) graph cumulant associated with the path graph with $3$ edges is: 
\begin{align}
    \kappa_{\threeline}^{ }(\mathcal{G}) = \mu_{\threeline}^{ }(\mathcal{G}) - 2\mu_{\twowedge}^{ }(\mathcal{G})\mu_{\oneedge}^{ }(\mathcal{G}) - \mu_{\twoparallel}^{ }(\mathcal{G})\mu_{\oneedge}^{ }(\mathcal{G}) + 2\mu_{\oneedge}^3(\mathcal{G}). 
    \label{Eq:GraphCumulant3Line}
\end{align}

\vspace{2pt}

\vspace{10pt}

\section{The Statistical Tests (How we Compare Graphs)}
\label{Sec:StatTest}
In this section, we describe the ways in which we ensured that the tests use the statistics $\mu_g^{ }$ and $\kappa_g^{ }$ in the same way, so as to compare graph cumulants on an equal footing with the corresponding subgraph densities/moments.  
In particular, both tests have the same structure (Section~\ref{ssec:TestStructure}), rely on the same assumptions (Section~\ref{ssec:Assumptions}), and have the same computational complexity (Section~\ref{ssec:TestComplexity}). 

\newpage
\subsection{Same Structure}
\label{ssec:TestStructure}
To compare graph moments and graph cumulants on the same footing, we use them as analogous inputs to the same simple statistical test:\\[-15pt]
\begin{enumerate}
    \item choose $r$, the maximum order of the subgraph statistics being considered;\\[-17pt] 
    \item for each sample, estimate its distribution using these subgraph statistics;\\[-17pt] 
    \item quantify the difference between distributions\\
    using a notion of distance for the space of these subgraph statistics.  \\[-15pt]
\end{enumerate}
In particular, we consider the graph moments/cumulants associated with \mbox{$\boldsymbol{g}_{r}^{\ExpLeftPar\textrm{c}\ExpRightPar}$}, the set of connected subgraphs with at most $r$ edges. 
To measure the difference between two samples of graphs $\mathbf{G}_{\shrinkA A}^{ }$ and $\mathbf{G}_{\shrinkB B}^{ }$, we use the squared Mahalanobis distance \citep{mahalanobis1936generalized} between their inferred moments/cumulants:
\begin{align}
\widehat{d_\mu^{2}}(\boldsymbol{G}_{\shrinkA A},\boldsymbol{G}_{\shrinkB B}) &= \big(\boldsymbol{\widehat{\mu}}_A^{ } - \boldsymbol{\widehat{\mu}}_B^{ }\big)^\top \big(\boldsymbol{\widehat{\Sigma}}_{A}^{\ExpLeftPar\mu\ExpRightPar} + \boldsymbol{\widehat{\Sigma}}_{B}^{\ExpLeftPar\mu\ExpRightPar}\big)^{-1} \big(\boldsymbol{\widehat{\mu}}_A^{ } - \boldsymbol{\widehat{\mu}}_B^{ }\big), \label{eq:MahaMom}\\
\widehat{d_\kappa^{2}}(\boldsymbol{G}_{\shrinkA A},\boldsymbol{G}_{\shrinkB B}) &= \big(\boldsymbol{\widehat{\kappa}}_A^{ } - \boldsymbol{\widehat{\kappa}}_B^{ }\big)^\top \big(\boldsymbol{\widehat{\Sigma}}_{A}^{\ExpLeftPar\kappa\ExpRightPar} + \boldsymbol{\widehat{\Sigma}}_{B}^{\ExpLeftPar\kappa\ExpRightPar}\big)^{-1} \big(\boldsymbol{\widehat{\kappa}}_A^{ } - \boldsymbol{\widehat{\kappa}}_B^{ }\big). \label{eq:MahaCum}
\end{align}
where \mbox{$\boldsymbol{\widehat{\mu}}_A^{ } = \boldsymbol{\widehat{\mu}}(\boldsymbol{G}_{\shrinkA A}^{ })$} is the vector of estimated moments of sample $\boldsymbol{G}_{\shrinkA A}^{ }$, and \mbox{$\boldsymbol{\widehat{\Sigma}}_B^{\ExpLeftPar\kappa\ExpRightPar}$} is the covariance estimate of the vector of estimated cumulants of sample $\boldsymbol{G}_{\shrinkB B}^{ }$, etc. 

In Section~\ref{Sec:StatTestMeanAndCov}, we describe how to compute these estimators. 

\subsection{Same Assumptions}
\label{ssec:Assumptions}
We assume that the $s$ observed graphs in a sample $\boldsymbol{G}$ were all obtained by subsampling $n$ nodes i.i.d.~from a single (much larger) ``graph'' $\mathcal{G}$. 
In other words, we assume that each graph distribution is generated by an underlying graphon \citep{borgs2008convergent, lovasz2012large}.  
This assumption allows for several notable simplifications (in particular, Equation~\ref{Eq:ProductToDisjointUnion}) without changing the primary message.

In addition, the graph distributions induced by a graphon are dense, 
and have empirical subgraph densities that converge to a Gaussian distribution as $n\rightarrow\infty$ \citep{bickel2011method}.  
As the empirical estimators of graph cumulants are linear combinations of the empirical subgraph densities (see Section~\ref{eq:UnbiasedCum}), 
they also limit to a Gaussian distribution.  
Moreover, as the number of graphs per sample becomes large ($s\rightarrow\infty$), the averages will also approach a Gaussian distribution by the central limit theorem (see also Appendix~\ref{Sec:PARE}).

In light of these Gaussian limits, 
the \mbox{well-established} Mahalanobis distance \citep{anderson1958introduction, reiser2001confidence} is appropriate for both tests considered in this paper.  
In Sections \ref{Sec:ResultsSimulations},~\ref{Sec:ResultsRealData}, and~\ref{Sec:ResultsChiSquared}, we show that (in contrast to moments) graph cumulants work well in practice, even when using this simple classical distance.

\subsection{Same Complexity}
\label{ssec:TestComplexity}
The computational complexity of both tests is determined by the complexity of counting the relevant subgraphs.
Computing the covariance matrices in Equations~\ref{eq:MahaMom} and~\ref{eq:MahaCum} requires counting connected subgraphs with at most $2r$ edges (see Section~\ref{SubSec:StatTestCov}).

For this paper, we use \mbox{$r=3$}.  
To count all but the complete graph on $4$ nodes ($K_4^{ }$), we follow an approach similar to that in \citet{maugis2016fast}.  
Beginning with the observed adjacency matrix, we require \mbox{$n$$\times $$n$} matrix multiplication and \mbox{entry-wise} operations (thereby taking $\mathcal{O}(n^\omega_{\vphantom{1}})$ time, where $\omega$ is the matrix multiplication exponent).  
For $K_4^{ }$, we require multiplying an \mbox{$n$$\times$$m$} and an \mbox{$m$$\times$$n$} matrix, where $m$ is the number of edges.

Our approach is not the best one can do.  
Efficiently counting subgraphs is a highly active area of research \citep{ribeiro2021survey}, 
with a variety of powerful algorithms \citep{ye2022lightning,pinar2017escape}, 
especially for certain (realistic) classes of graphs \citep{bera2019linear, chiba1985arboricity,bressan2021faster}, and for particular subgraphs (e.g., triangles \citep{gui2019fast}).
Asymptotics aside, substantial acceleration can be obtained through parallel computation \citep{dodeja2022parsec, biswas2022massively}, and more recently through the use of machine learning techniques \citep{zhao2021learned,liu2020neural}.

\section{How to Estimate the Statistics (What we Compute)}
\label{Sec:StatTestMeanAndCov}
In Section~\ref{Sec:StatTestMean}, we describe unbiased estimators for the graph moments and cumulants, highlighting an important relationship between subgraph densities in the process (Equation~\ref{Eq:ProductToDisjointUnion}).  
In Section~\ref{SubSec:StatTestCov}, we build on this idea to describe the typical fluctuations of such sample statistics in terms of their covariance matrices. 

\subsection{Obtaining Unbiased Estimators (Getting the Mean)} 
\label{Sec:StatTestMean}
\subsubsection{For Graph Moments (Simple Substitution)} 
\label{SubSec:StatTestMean_Moments}
For a graphon model, graph moments are preserved under node subsampling:
\begin{equation}
    \big\langle \mu_g^{ }(G) \big\rangle_{G\sim\mathcal{G}}^{ } = \mu_g^{ }(\mathcal{G}).\label{eq:momentsPreserved}
\end{equation}
Thus, the empirical graph moments $\boldsymbol{\widehat{\mu}}(G)$ are themselves unbiased estimators of the moments of the distribution $\boldsymbol{\mu}(\mathcal{G})$ from which they were sampled.  
In particular, for a sample of $s$ graphs: 
\begin{equation}
\widehat{\mu}_{g}^{ }(\boldsymbol{G}) = \frac{1}{s}\sum_{i=1}^s \mu_{g}^{ }(G_i^{ }).\label{eq:avgMoments}
\end{equation}

\subsubsection{For Graph Cumulants (Slightly Subtle)} 
\label{SubSec:StatTestMean_Cumulants}
To obtain the analogous estimators for graph cumulants, we must be slightly more careful, 
as products of graph moments are not preserved in expectation under node subsampling.  
Fortunately, for graphs sampled from a graphon, products of graph moments are equal to the moment of their disjoint union\footnote{Intuitively, nodes in disjoint components have nothing to do with each other, so the presence of edges in any component are independent from the presence of edges in the other.} \citep{lovasz2012large, maugis2020testing}: 
\begin{align}
    \mu_g^{ }(\mathcal{G}) \mu_{g'}^{ }(\mathcal{G}) = \mu_{g\cup g'}^{ }(\mathcal{G}). \label{Eq:ProductToDisjointUnion}
\end{align}
For example, using this relation in the expression for the cumulant associated with the path graph with $3$ edges (Equation~\ref{Eq:GraphCumulant3Line}) results in its unbiased estimator: 
\begin{align}
    \widehat{\kappa}_{\threeline}^{ }(G) = \mu_{\threeline}^{ }(G) - 2\mu_{\threeedgewedge}^{ }(G) + \mu_{\threeparallel}(G). \label{Eq:ThreelineUnbiased}
\end{align}

For a sample of $s$ graphs, the estimated cumulants of the underlying distribution are the average of the unbiased estimators of each individual graph: 
\begin{align}
   \widehat{\kappa}_g^{ }(\boldsymbol{G}) = \frac{1}{s}\sum_{i=1}^s\widehat{\kappa}_g^{ }(G_i^{ }).  \label{eq:UnbiasedCum}
\end{align}

\subsection{Analytically Computing Significance (Getting the Covariance)}
\label{SubSec:StatTestCov}
Consider sampling graphs from a distribution $\mathcal{G}$ with known graph moments.  
The covariance between their observed graph moments is 
\begin{align}
    \text{Cov}\big(\mu_g^{ }(G),\mu_{g'}^{ }(G)\big) = \big\langle\mu_g^{ }(G) \mu_{g'}^{ }(G)\big\rangle - \big\langle\mu_g^{ }(G)\big\rangle\big\langle\mu_{g'}^{ }(G)\big\rangle. \label{Eq:Covariance}
\end{align} 
The last term is trivial; as graph moments are preserved under node subsampling (Equation~\ref{eq:momentsPreserved}), 
the expectations of the empirical moments are exactly the moments of the underlying distribution.  
The first term, however, is the expectation of a \textit{product} of graph moments. 

Equation~\ref{Eq:ProductToDisjointUnion} applies to a single graphon $\mathcal{G}$.  
The analogous expression for the moments of a single finite graph $G$ 
requires considering all the ways that the nodes of the subgraphs could overlap \citep{maugis2020testing}.  
Thus, estimating the covariance of the empirical graph moments/cumulants up to order $r$ requires the estimation of the distribution's graph moments/cumulants up to order $2r$ (as stated in Section~\ref{ssec:TestComplexity}).
For example, in terms of injective homomorphism counts, we have the following expression: 
\begin{align}
    c_{\twowedge}^{ }(G) c_{\oneedge}^{ }(G) = 4c_{\twowedge}^{ }(G) + 2c_{\threetriangle}^{ }(G) + 2c_{\threeclaw}^{ }(G) + 4c_{\threeline}^{ }(G) + c_{\threeedgewedge}^{ }(G). \label{eq:CountProductExample}
\end{align}
The $4c_{\twowedge}^{ }$ term corresponds to the four ways that the two nodes of the edge can be placed to overlap with the wedge ($\WedgeBig$), the $2c_{\threetriangle}^{ }$ term corresponds to the two ways those nodes can be placed to turn a wedge into a triangle, and so on.
Converting from counts to moments and taking the expectation allows one to express Equation~\ref{Eq:Covariance} in terms of moments of the distribution.   

As the $s$ graphs in each sample are i.i.d.~and have the same number of nodes, the covariance of the sample graph moments is 
\begin{align}
   \boldsymbol{\widehat{\Sigma}}_{\vphantom{1}}^{\ExpLeftPar\mu\ExpRightPar}(\boldsymbol{G}) = \frac{1}{s}\sum_{i=1}^s\boldsymbol{\widehat{\Sigma}}_{\vphantom{1}}^{\ExpLeftPar\mu\ExpRightPar}(G_i^{ }).  \label{eq:MomCov}
\end{align}

As the unbiased graph cumulants are linear combinations of the sample graph moments, obtaining their covariance follows \textit{mutatis mutandis}:  
\begin{align}
   \boldsymbol{\widehat{\Sigma}}_{\vphantom{1}}^{\ExpLeftPar\kappa\ExpRightPar}(\boldsymbol{G}) = \frac{1}{s}\sum_{i=1}^s\boldsymbol{\widehat{\Sigma}}_{\vphantom{1}}^{\ExpLeftPar\kappa\ExpRightPar}(G_i^{ }).  \label{eq:CumCov}
\end{align}

\newpage
\section{A Controlled Competition (Between the Tests)} 
\label{Sec:ResultsSimulations}
We first describe the models we use to generate the synthetic data. 
We then describe how we summarize the quality of the statistical tests, with illustrations for some representative simulations. 
Asymptotic results as \mbox{$s\rightarrow\infty$} are presented in Appendix~\ref{Sec:PARE}.

\subsection{Synthetic Data (A Pair of \mbox{two-by-two} SBMs)} 
\label{SubSec:SyntheticExperiments}
We consider two graph distributions: one with heterogeneous degree distribution (but no community structure), and the other with community structure (but homogeneous degree distribution).  
Both distributions are Stochastic Block Models (SBMs) \citep{holland1983stochastic} with two \mbox{equal-sized} communities and expected edge density $\rho$.

The heterogeneous SBM is parameterized by $\varepsilon_h^{ }$, with connectivity matrix: 
\begin{align}
\mat{B}_h^{ } = \rho \times \begin{bmatrix}
1+\varepsilon_h^{ } & 1\\
1 & 1-\varepsilon_h^{ }
\end{bmatrix},
\label{eq:B2het}
\end{align}
where \mbox{$\varepsilon_h^{ } = 0$} gives uniform connection probability between all pairs of nodes, 
and \mbox{$\varepsilon_h^{ } = 1$} gives zero connection probability within one of the two communities.  

The assortative SBM is parameterized by $\varepsilon_a^{ }$, with connectivity matrix: 
\begin{align}
\mat{B}_a^{ } = \rho \times \begin{bmatrix}
1+\varepsilon_a^{ } & 1-\varepsilon_a^{ }\\
1-\varepsilon_a^{ } & 1+\varepsilon_a^{ }
\end{bmatrix},
\label{eq:B2assort}
\end{align}
where \mbox{$\varepsilon_a^{ } = 0$} again gives uniform connection probability between all pairs of nodes, 
and \mbox{$\varepsilon_a^{ } = 1$} gives zero connection probability between the two communities. 

After fixing two such SBMs, each instantiation of a \mbox{two-sample} test involves ``flipping two coins'' to decide which distributions each of the two samples will come from.  
That is, the instantiations are split evenly between ``different distributions'' and ``same distribution'', with the latter split evenly between the two SBMs.

\subsection{ROC and AUC Curves (How we Compare Tests)}
\label{SubSec:ROCCurves}
\begin{figure}[H]
\begin{center}
\centerline{\includegraphics[width=0.55\columnwidth]{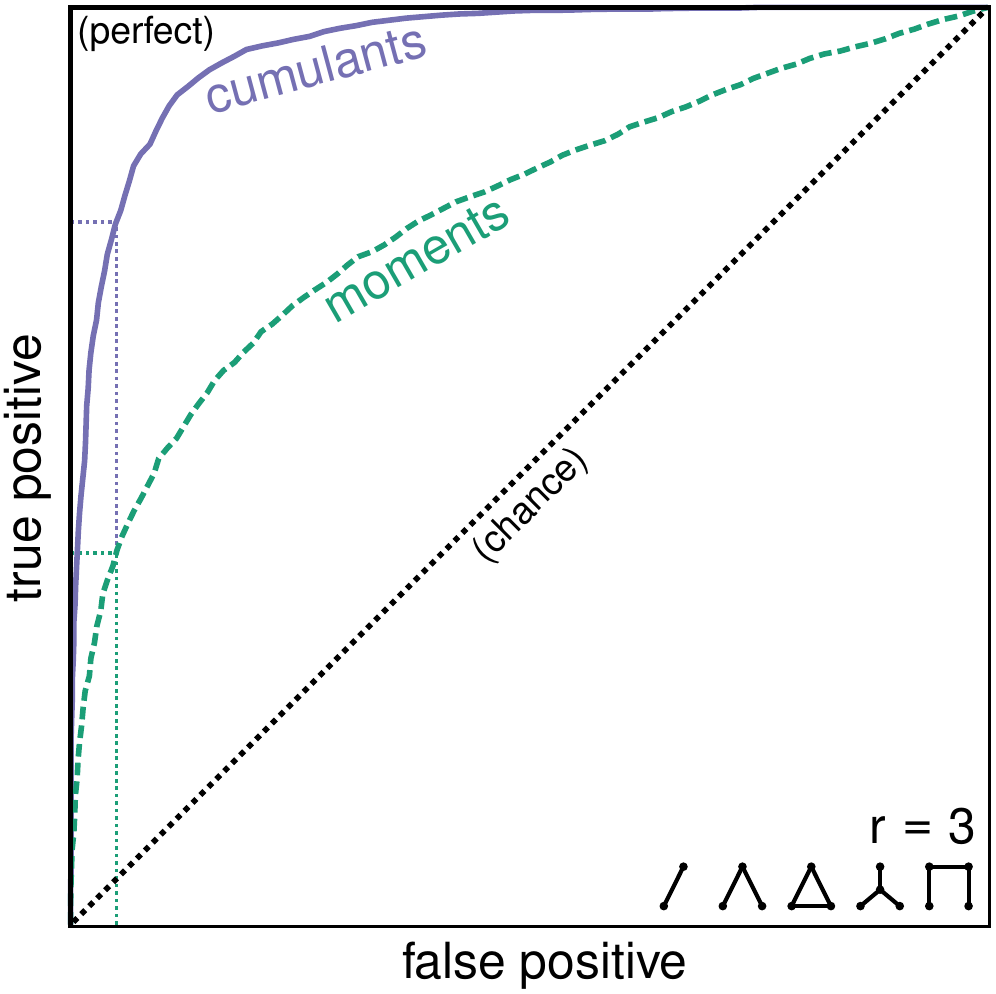}} 
\vspace{-2pt}
\caption{
\mbox{\textbf{The statistical power for all possible error rates.} } \mbox{\hspace{0.19\textwidth} }
Known as an ROC curve, this plot visualizes the possible rates of false positives (Type \textrm{I} errors) and false negatives (Type \textrm{II} errors) of a binary classification method.  
Often, one specifies a maximum rate of false positives, commonly known as $\alpha$.   
The vertical dotted lines in the lower left illustrate the canonical choice of \mbox{$\alpha=0.05$}, and   
the point where this vertical line meets the ROC curves gives the corresponding rate of true positives (i.e., the statistical power, or \mbox{$1-(\text{Type II error rate})$}).  
An ROC curve displays the results for all possible values of $\alpha$: 
random guesses result in a line along the diagonal, 
while perfect answers result in a line that hugs the upper left.  
For this plot, both statistical tests \mbox{(\textcolor{momentscolor}{\textbf{moments}} and \textcolor{cumulantscolor}{\textbf{cumulants}})} 
use the counts of connected subgraphs with up to \mbox{$r=3$} edges to distinguish between samples from two graph distributions: a heterogeneous SBM (Equation~\ref{eq:B2het} with \mbox{$\varepsilon_h^{ }=\tfrac{1}{16}$}), and an assortative SBM (Equation~\ref{eq:B2assort} with \mbox{$\varepsilon_a^{ }=\tfrac{1}{16}$}), both with density \mbox{$\rho=\tfrac{1}{2}$}.  
Each sample contains \mbox{$s=4$} graphs, and all graphs have \mbox{$n=256$} nodes.
}
\label{Fig:ROCExample}
\end{center}
\end{figure}

After computing the squared Mahalanobis distance between a pair of samples (Equations~\ref{eq:MahaMom} and~\ref{eq:MahaCum}), we use a threshold to classify them as coming from the ``same distribution'' or ``different distributions.''  
Each choice of threshold induces: a rate of false positives (incorrectly concluding that the distributions are different), and a rate of true positives (correctly concluding that the distributions are different).  
All possible threshold choices are summarized in a Receiver Operating Characteristic (ROC) curve (see Figure~\ref{Fig:ROCExample}).  
\newpage

\begin{figure}[H]
\begin{center}
\centerline{\includegraphics[width=0.9\columnwidth]{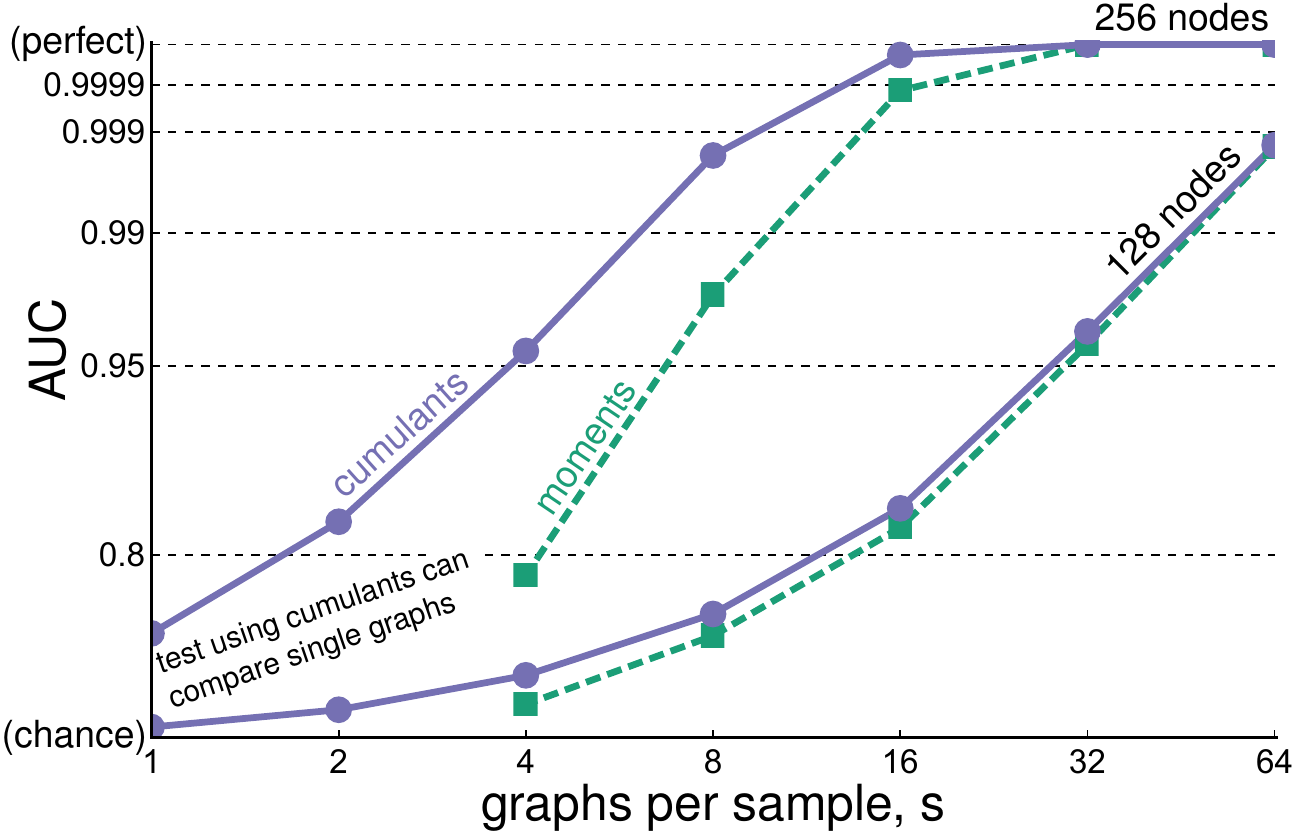}} 
\vspace{-2pt}
\caption{\mbox{\textbf{Using graph cumulants outperforms using graph moments,} } 
\mbox{\textbf{particularly when the number of graphs per sample is small.} } \mbox{\hspace{0.07\textwidth}} 
We performed the comparison in Figure~\ref{Fig:ROCExample} for a range of graphs per sample $s$, 
for graphs with \mbox{$n=128$} nodes and graphs with \mbox{$n=256$} nodes. 
And we summarized the resulting ROCs in terms of the the Area Under the Curve (AUC).   
The (significantly distorted) vertical axis ranges from $0.5$ (chance) to $1.0$ (perfect).  
The test using moments (with at most \mbox{$r=3$} edges) fails to give an answer for \mbox{$s<4$} (see Appendix~\ref{Appendix:CovS}).  
In contrast, the test using cumulants works even when there is only a \textit{single} observed graph in each sample (see Section~\ref{Sec:Discussion}).  
}
\label{Fig:SBM2NandS}   
\end{center}
\end{figure}

To compare ROC curves, we summarize them with a single scalar value, viz., the Area Under the Curve (AUC).  
Figure~\ref{Fig:SBM2NandS} compares the AUC for graphs with \mbox{$n=128$} and \mbox{$n=256$} nodes over a range of sample sizes.  
The use of graph cumulants consistently results in greater statistical power, especially when the number of graphs per sample is small (including the case of a \textit{single}  graph per sample, for which the test using moments is not applicable).

One might wonder if these results are sensitive to the model parameters $\varepsilon_h^{ }$ and $\varepsilon_a^{ }$.  
While the values used to create the figures were judiciously chosen (e.g., such that the AUC spans the range from chance to perfect), the qualitative conclusions remain robust for (essentially) any parameter choice. 
In the next section, we show that the use of graph cumulants is similarly promising for applications to more realistic networks.  

\vspace{2pt}

\section{Studying the Tests in the Wild (Application to Real Networks)}  
\label{Sec:ResultsRealData} 
Most \mbox{real-world} networks look rather different than the simulated graphs in the previous section.  
In this section, we compare the two tests using graphs sampled from more realistic graph distributions.  
In particular, 
we use gene interaction networks from four different species: Mouse, Rat, Human, and Arabidopsis (a small flowering plant related to cabbage and mustard), all curated by the FunCoup repository \citep{PERSSON2021166835}.

\begin{figure}[H]
\begin{center}
\centerline{\includegraphics[width=0.97\columnwidth]{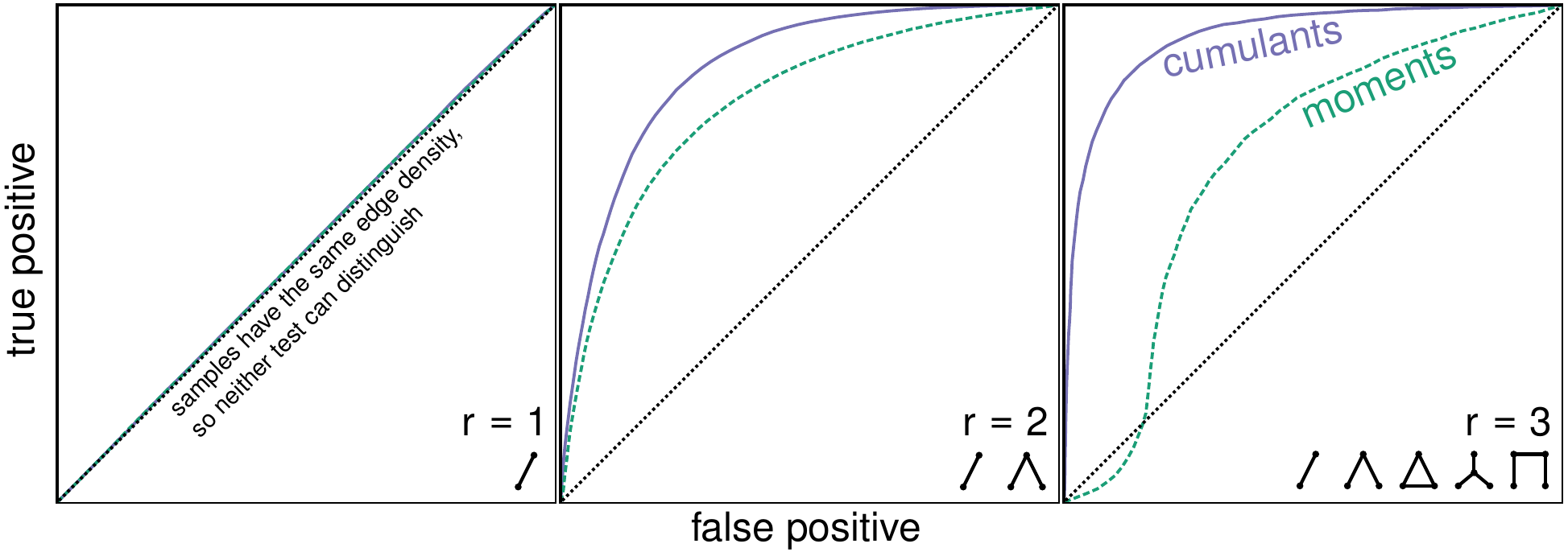}}
\vspace{-2pt}
\caption{
\mbox{\textbf{Using more subgraphs increases accuracy for cumulants, }} \hspace{\textwidth}
\mbox{\textbf{but causes overfitting for moments. }} \mbox{\hspace{\textwidth} \hspace{\textwidth}}
We use the two tests to compare the genetic interaction networks of Arabidopsis (\mbox{$\sim$$1.3\!\times\!10^4$} nodes, \mbox{$\sim$$7.9\!\times\!10^5$} edges) and Mouse 
(\mbox{$\sim$$1.4\!\times\!10^4$} nodes, \mbox{$\sim$$9.2\!\times\!10^5$} edges) (data from \citet{PERSSON2021166835}).  
As these networks were adjusted to have the same edge density (see Section~\ref{Sec:ResultsRealData}), 
neither \mbox{$r=1$} test (left) can distinguish between them.  
The benefit of using cumulants can be seen for the tests using more subgraphs.  
When including subgraphs with \mbox{$r=2$} edges (middle), the test using cumulants consistently outperforms the test using moments.  
When including subgraphs with \mbox{$r=3$} edges (right), the test using cumulants continues to improve, while the test using moments performs worse.   
Each sample contained \mbox{$s=4$} graphs, and each graph had \mbox{$n\sim256$} nodes (see Section~\ref{Sec:ResultsRealData}).
}
\label{Fig:RealDataComparisonR}
\end{center}
\end{figure}

Although these networks have relatively similar edge densities, 
the differences are already enough for either test to easily distinguish samples from them using only the edge subgraph.  
As the differences between moments and cumulants only become apparent for subgraphs with more than one edge, %
we adjusted the networks being compared so they have the same edge density. 
This was done by removing the \mbox{lowest-weight} edges from the network with larger edge density until it had the same density as the other.\footnote{If one wishes to test if two distributions with different sparsities are otherwise the same, one could scale all subgraph densities by the edge density to the power of the number of edges in the subgraph.} 
The effect of this adjustment can be seen in Figure~\ref{Fig:RealDataComparisonR} (left); 
both tests using \mbox{$r=1$} edge can no longer distinguish between the samples.

Each sample contains $s$ graphs, all obtained from one of these adjusted networks by subsampling its nodes.   
Specifically, each graph $G$ in a sample contains $n$ nodes in expectation 
(selected from the $N$ nodes in the adjusted network independently with probability $\tfrac{n}{N}$), 
and nodes in $G$ are connected by an edge if (and only if) their corresponding nodes in the adjusted network are connected by an edge. 
That is, $G$ is an induced subgraph of the adjusted network.

\begin{figure}[H]
\begin{center}
\centerline{\includegraphics[width=0.97\columnwidth]{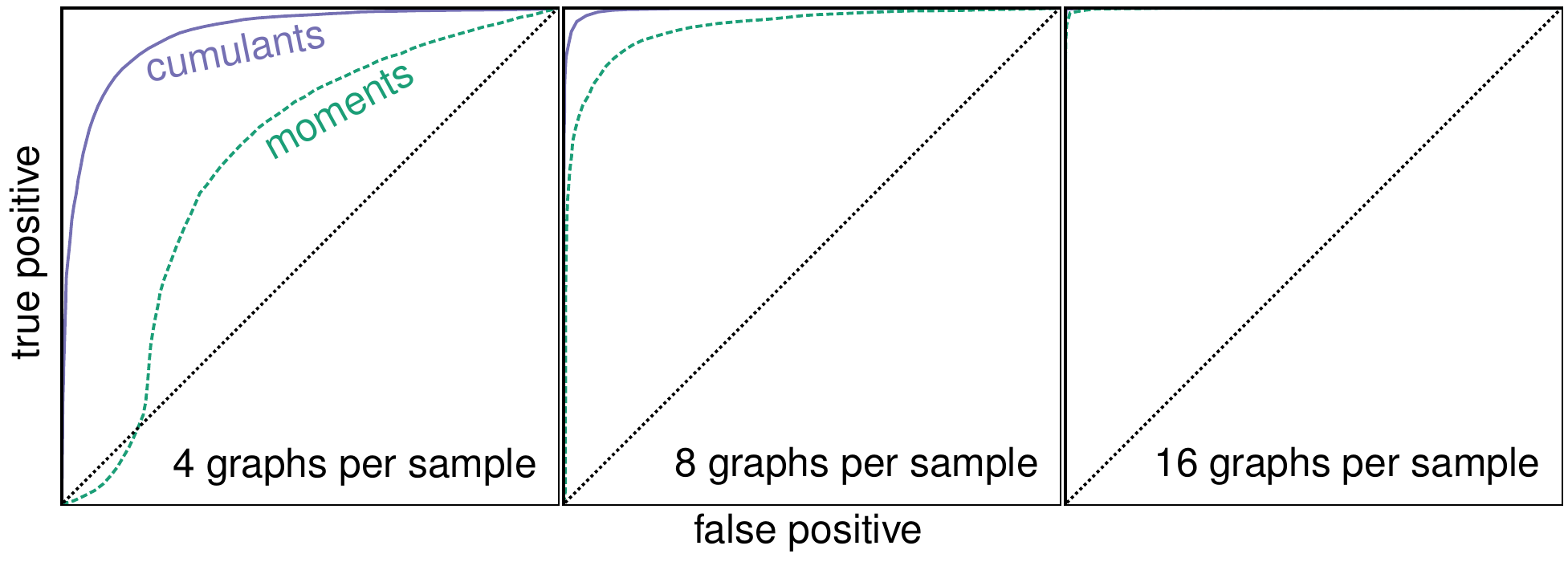}}
\vspace{-2pt}
\caption{
\mbox{\textbf{Cumulants continue to consistently outperform moments in real data,}} \hspace{\textwidth} 
\mbox{\textbf{especially when the number of graphs per sample is small.}} \hspace{\textwidth} 
We compare the genetic interaction networks of Human (\mbox{$\sim$$1.6\!\times\!10^4$} nodes, \mbox{$\sim$$13.9\!\times\!10^5$} edges) and Rat (\mbox{$\sim$$1.1\!\times\!10^4$} nodes, \mbox{$\sim$$9.0\!\times\!10^5$} edges).  
The samples are generated with the same method as in Figure~\ref{Fig:RealDataComparisonR}, and the tests use connected subgraphs with up to \mbox{$r=3$} edges. 
}
\label{Fig:RealDataComparisonS}
\end{center}
\end{figure}

Figures~\ref{Fig:RealDataComparisonR} and~\ref{Fig:RealDataComparisonS} show that using graph cumulants to discriminate between these biological networks outperforms the analogous test using graph moments.

\vspace{2pt}

\vspace{2pt}

\vspace{10pt}
\section{Why Cumulants do Better}
\label{Sec:ResultsChiSquared}
Why does using graph cumulants consistently provide better statistical power?  
After all, unbiased graph cumulants are simply linear combinations of graph moments!  (Section~\ref{SubSec:StatTestMean_Cumulants})  
Loosely, the reason is that graph distributions ``look more Gaussian'' when represented in terms of cumulants (as compared to moments).

Both tests measure differences between distributions using the squared Mahalanobis distance (Equations~\ref{eq:MahaMom} and~\ref{eq:MahaCum}). 
As this measure depends only on the mean and covariance, there is a tacit assumption that the sample statistics are \mbox{well-characterized} by a multivariate Gaussian \citep{rao1992information}.  

Indeed, if the subgraph statistics \textit{were} precisely Gaussian, 
and their covariance known \textit{exactly},  
then the squared Mahalanobis distance between pairs of samples from the same distribution would in fact follow a \mbox{$\chi^2_{\vphantom{1}}$} distribution.  
However, as the sample statistics are \textit{not} precisely Gaussian, and their covariance must be \textit{estimated}, one should indeed expect some deviation from this limiting distribution.

\begin{figure}[H]
\begin{center}
\centerline{\includegraphics[width=0.99\columnwidth]{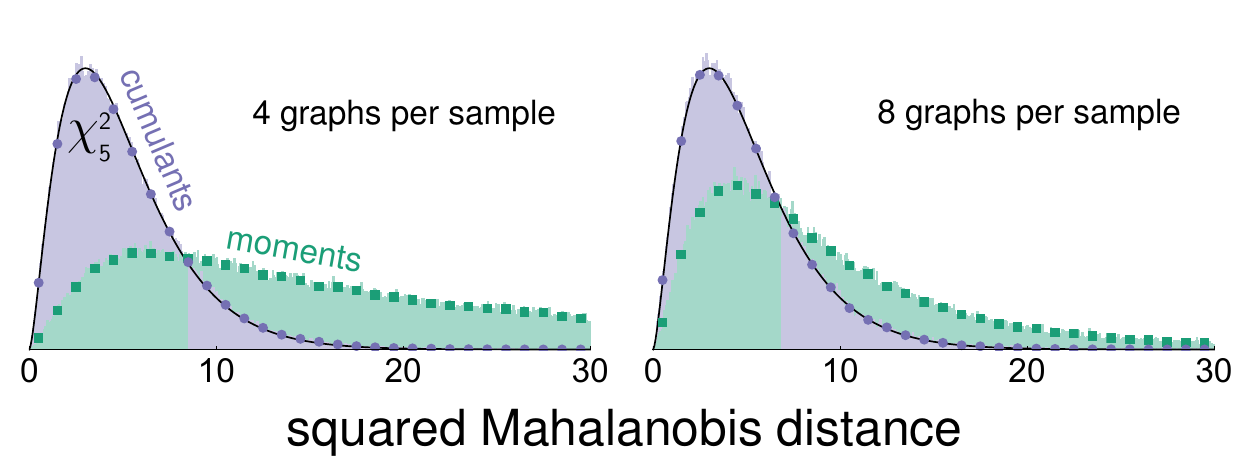}} 
\vspace{-2pt}
\caption{
\mbox{\textbf{The test statistic based on graph cumulants remains nearly $\chi^2_{ }$,}} \hspace{\textwidth} 
\mbox{\textbf{even when only few graphs are observed.}} \mbox{\hspace{\textwidth} \hspace{\textwidth}}
If a test statistic follows a known distribution under the null hypothesis, one can control the false positive rate by choosing an appropriate threshold.  
Indeed, for a large number of graphs per sample \mbox{$s\rightarrow\infty$}, both tests converge to a \mbox{$\chi^2_{ }$ distribution}.  
As the number of graphs per sample decreases, this becomes a poor approximation for the test using moments.  
In contrast, the approximation remains strikingly robust for the test using cumulants.  
Colored histograms are the empirical distributions of the test statistics (squared Mahalanobis distance) for simulations with the same parameters as in Figure~\ref{Fig:ROCExample}, and black curves are the limiting \mbox{$\chi^2_{ }$ distribution} (with $5$ degrees of freedom).
}
\label{Fig:ChiSquared}
\end{center}
\end{figure}

As illustrated in Figure~\ref{Fig:ChiSquared}, 
when the graphs are sampled from the same distribution, 
the test using moments results in a distribution of squared Mahalanobis distances with \mbox{overly-heavy} tails. 
Whereas the distribution for the test using cumulants is \mbox{well-characterized} by the appropriate \mbox{$\chi^2_{\vphantom{1}}$} distribution, 
thereby allowing one to control the false positive rate. 

\newpage

\vspace{2pt}

\section{The Main Message (Discussion)}
\label{Sec:Discussion}
Perhaps the most salient advantage of using graph cumulants (instead of the subgraph densities themselves) is 
the ability to compare distributions even when observing only a \textit{single} graph from each (e.g., Figure~\ref{Fig:SBM2NandS}).  

At first glance, it may seem strange that one could make inferences about a distribution from a sample containing only a single ``observation.''  
Indeed, this certainly does not work for \mbox{scalar-valued} random variables --- a single observation provides no information about the spread of its underlying distribution.  

Essentially, this difference arises because, in graphs, the data reside in the edges, but sampling is applied \mbox{node-wise}. 
Notationally, this manifests itself in our need to specify both $n$ (the number of nodes per graph) and $s$ (the number of graphs per sample), whereas the ``quantity'' of i.i.d.~scalar data is specified by only the sample size. 

Thus, there are two relevant limits: we may ask about the distributions of these statistics as either $n$ or $s$ become sufficiently large.  
As \mbox{$s\rightarrow\infty$} (with fixed $n$), both methods are asymptotically normal by the central limit theorem.  
In Appendix~\ref{Sec:PARE}, we show that using graph cumulants remains statistically advantageous even in this ``classical'' limit.
As \mbox{$n\rightarrow\infty$} (with fixed $s$),
graph cumulants appear to more properly exploit the multiplicity of the data \textit{within} each graph, allowing for inference even when \mbox{$s=1$} (see Appendix~\ref{Appendix:CovS}).

In a sense, graph cumulants provide a more ``natural'' set of coordinates than the subgraph densities themselves.  
For example, the edge density $\mu_{\oneedge}^{ }$ is strongly correlated with the wedge moment $\mu_{\twowedge}^{ }$.  
In contrast, $\kappa_{\twowedge}^{ }$ ``takes into account'' the \mbox{lower-order} effect of edge density, rendering it ``more orthogonal'' to $\mu_{\oneedge}^{ }$.  
In fact, 
for an \mbox{Erd\H{o}s-R\'enyi} distribution, 
the covariance between the edge density and all other graph cumulants $\widehat{\kappa}_{g}^{ }$ is precisely zero (see proof in Appendix~\ref{Sec:ERCumCov}).

In practice, the covariance estimates for cumulants are more robust, leading to impressively accurate agreement with the classical $\chi^2_{\vphantom{1}}$ distribution.
This allows one to convert these statistics into precise probabilistic statements (e.g.: ``What is the likelihood that two graphs were generated by different graphons?'').  
This notion of ``graphonic similarity'' between any pair of graphs is a remarkably general tool.

\section{Promising Sequels (Future Directions)} 
\label{Sec:PostScript}

Towards the goal of estimating the relevance of various subgraphs/motifs in an observed network, 
recent works \citep{bhattacharya2022motif, zhang2022edgeworth} have provided  \mbox{higher-order} characterizations of the sampling distributions of (appropriately rescaled) estimators of subgraph densities for various regimes and model assumptions. 
Given the success of graph cumulants compared to the bare densities/moments in determining network similarity, 
a natural step to help in this endeavor is to provide 
such characterizations for the estimators of graph cumulants (e.g., by generalizing the result in Appendix~\ref{Sec:ERCumCov} to arbitrary graphons and pairs of subgraphs).

It might be argued that intelligent feature engineering has been rendered moot by powerful supercomputers fitting deep neural networks to big data.  
Nonetheless, some of the most notable breakthroughs in machine learning have been the result of architectures with cleverly engineered priors (e.g., convolutional neural networks for computer vision tasks \citep{lecun1998gradient, khan2018guide}, and transformers for natural language processing tasks \citep{vaswani2017attention, chan2022data}). 
Like the edge and orientation detection cells in the first layers of visual processing \citep{hubel1959receptive, gabbiani2017mathematics}, 
graph cumulants offer a principled first layer of contrastive features when looking at the relationships between edges in networks.

As a coda,
we highlight a compelling analogy between the conversion between induced subgraph densities and 
injective homomorphism subgraph densities 
and the conversion between injective homomorphism 
subgraph densities (i.e., graph moments) 
and their corresponding graph cumulants.  
The former applies a M\"obius transform to the poset induced by the inclusion of edges in a subgraph with a fixed number of nodes.  
The latter also applies a M\"obius transform, though with respect to the partial order induced by the partitions of an arrangement of a fixed number of edges.

This suggests a general framework for the ``cumulantification'' of other types of combinatorial structures.  
Indeed, extensions to hypergraphs and directed networks 
were described in the Appendices of \citet{Gunderson2019Introducing}, 
though the applications appear to be even more general --- 
any combinatorial structure admitting a notion of ``reduction'' or ``partitioning'' naturally induces a corresponding poset over its substructures, offering an avenue for obtaining similarly useful statistics.

\newpage

\acks{In addition to \mbox{Tina~Eliassi-Rad} and anonymous reviewers for their comments and patience,\\ the authors are grateful for the sage wisdom of \mbox{Ashlyn~Maria~Bravo~Gundermsdorff}.  
}

\appendix

\section{\mbox{The Limit of Many Observed Graphs} \mbox{ (Asymptotic Relative Efficiency of the Two Tests)}} 
\label{Sec:PARE}
In Section~\ref{Sec:ResultsChiSquared}, 
our simulations show that the test statistic based on graph cumulants closely approximates a $\chi^2_{\vphantom{1}}$ distribution under the null hypothesis even for a small number of observed graphs $s$, whereas the analogous test statistic based on graph moments does not.
However, both statistics indeed do converge to a \mbox{$\chi^2_{\vphantom{1}}$ distribution} in the limit of many observed graphs \mbox{$s\rightarrow\infty$} \citep{maugis2020testing}. 

In this section, we compare the statistical power of the two tests in this limit of large sample size, showing that using cumulants still outperforms using moments.

\subsection{The Model}
\label{ssec:paremodel}
To this end, we consider the large sample size limit \mbox{$s\rightarrow\infty$} as the distributions become increasingly similar \mbox{$\mathcal{G}_1^{ } \rightarrow \mathcal{G}_0^{ }$}.  
We represent these distributions as stochastic block models (SBMs): 
a probability vector $\boldsymbol{\pi}$, giving the expected fractions of nodes contained in each community,  
and a connectivity matrix $\mat{B}$, giving the probability of an edge between two nodes based on their assigned communities.

Let $\mathcal{G}_1^{ }$ be a perturbation of the form \mbox{$(\boldsymbol{\pi}_1^{ },\mat{B}_1^{ }) = (\boldsymbol{\pi}_0^{ },\mat{B}_0^{ }) + \sqrt{\gamma/s}\,(\deltapi,\deltab)$}.    
The \mbox{$1/\sqrt{s}$} scaling of this perturbation as \mbox{$s\rightarrow\infty$} 
is such that, 
given some maximum allowable error rates of false positives $\alpha$ and false negatives $\beta$ for a test, 
there exists a critical value \mbox{$\gamma_*^{ }$} above which these desiderata are achievable.

For two different tests, the ratio of their $\gamma_*^{ }$ is known as the Pitman asymptotic relative efficiency ($\pare$) \citep{pitman1949notes}.
As the two tests we are comparing have the same \mbox{$s\rightarrow\infty$} limiting distribution $\chi_5^2$ (as we are considering $r=3$), taking this ratio removes the dependence on $\alpha$ and $\beta$.  
Thus, the $\pare$ is given by the asymptotic ratio of the squared Mahalanobis distances, and depends on two choices: the distribution \mbox{$(\boldsymbol{\pi}_0^{ },\mat{B}_0^{ })$}, and the perturbation \mbox{$(\deltapi,\deltab)$}.

For \mbox{$(\boldsymbol{\pi}_0^{ },\mat{B}_0^{ })$}, we choose a model that allow us to independently control 
assortativity and (degree) heterogeneity.  
Specifically, we blend the two previously mentioned SBMs (Equations~\ref{eq:B2het} and~\ref{eq:B2assort}), using the Kronecker product of their connectivity matrices \mbox{$\mat{B}_h^{ }$ and $\mat{B}_a^{ }$}, i.e., 
\begin{align}
\mat{B}_0^{ } = \rho \times \begin{bmatrix}
\big(1+\varepsilon_h^{ }\big)\big(1+\varepsilon_a^{ }\big) & \big(1+\varepsilon_h^{ }\big)\big(1-\varepsilon_a^{ }\big) & \big(1+\varepsilon_a^{ }\big) & \big(1-\varepsilon_a^{ }\big)\\[2pt]
\big(1+\varepsilon_h^{ }\big)\big(1-\varepsilon_a^{ }\big) & \big(1+\varepsilon_h^{ }\big)\big(1+\varepsilon_a^{ }\big) & \big(1-\varepsilon_a^{ }\big) & \big(1+\varepsilon_a^{ }\big)\\[2pt]
\big(1+\varepsilon_a^{ }\big) & \big(1-\varepsilon_a^{ }\big) & \big(1-\varepsilon_h^{ }\big)\big(1+\varepsilon_a^{ }\big) & \big(1-\varepsilon_h^{ }\big)\big(1-\varepsilon_a^{ }\big)\\[2pt]
\big(1-\varepsilon_a^{ }\big) & \big(1+\varepsilon_a^{ }\big) & \big(1-\varepsilon_h^{ }\big)\big(1-\varepsilon_a^{ }\big) & \big(1-\varepsilon_h^{ }\big)\big(1+\varepsilon_a^{ }\big)
\end{bmatrix}, 
\label{eq:BforPARE}
\end{align}
and $\boldsymbol{\pi}_0^{ }$ is uniform over the \mbox{$k=4$} communities.

For \mbox{$(\deltapi,\deltab)$}, 
we choose random perturbations to be Gaussian with covariance proportional to typical fluctuations of realizations of this model \mbox{$(\boldsymbol{\pi}_0^{ },\mat{B}_0^{ })$} for graphs with $n$ nodes.  
Thus, $\deltapi$ has zero mean, with covariance given by the corresponding multinomial distribution: 
\begin{align}
\label{Eq:DeltaPi}
\text{Var}\big(\deltapi_{i}^{ }\big) = \frac{(\boldsymbol{\pi}_0^{ })_i^{ }\big(1-(\boldsymbol{\pi}_0^{ })_i^{ }\big)}{n}, 
\quad\text{Cov}\big(\deltapi_{i}^{ },\deltapi_{j}^{ }\big) = -\frac{(\boldsymbol{\pi}_0^{ })_i^{ }(\boldsymbol{\pi}_0^{ })_j^{ }}{n}.
\end{align}
Likewise, $\delta\mat{B}$ has zero mean, with variance given by: 
\begin{align}
\label{Eq:DeltaB}
\text{Var}\big(\delta\!B_{i\kern-0.5pt i}^{ }\big) = \frac{(\mat{B}_0^{ })_{i\kern-0.5pti}^{ }\big(1-(\mat{B}_0^{ })_{i\kern-0.5pt i}^{ }\big)}{{\tbinom{n}{2}}(\boldsymbol{\pi}_0^{ })_i^{2}}, 
\quad\text{Var}\big(\delta\!B_{i\!j}^{ } = \delta\!B_{j\kern-0.5pt i}^{ }\big) = \frac{(\mat{B}_0^{ })_{i\!j}^{ }\big(1-(\mat{B}_0^{ })_{i\!j}^{ }\big)}{2{\tbinom{n}{2}}(\boldsymbol{\pi}_0^{ })_i^{ }(\boldsymbol{\pi}_0^{ })_j^{ }}.
\end{align}

\subsection{The Computation}
\label{ssec:parecomputation}
To evaluate the $\pare$, 
we 
construct a symmetric matrix $\mat{M}$ that yields the squared Mahalanobis distance of a random perturbation by evaluating its quadratic form with a vector $\boldsymbol{\eta}$ containing i.i.d.~normal entries.  

We first obtain the graph moments up to order \mbox{$2r=6$} of the distribution defined by \mbox{$(\boldsymbol{\pi}_{0}^{ },\mat{B}_0^{ })$}, and use these to compute the covariance of sample moments up to order \mbox{$r=3$} (for graphs with $n$ nodes).  
We also obtain the derivatives 
$\mat{J}_{\partial{\mu}/\partial {(\deltapi,\deltab)}}$ 
of the moments up to order \mbox{$r=3$} with respect to the parameters defining \mbox{$(\deltapi,\deltab)$}.
We then multiply both sides of the inverse of the covariance matrix with the  matrix of partial derivatives: 
\begin{align}
    \big(\mat{J}_{\partial{\mu}/\partial {(\deltapi,\deltab)}}\big)^\top_{\vphantom{1}}
\big(\mat{\Sigma}_{\vphantom{1}}^{\ExpLeftPar\mu\ExpRightPar}\big)^{-1}_{\vphantom{1}} \big(\mat{J}_{\partial{\mu}/\partial {(\deltapi,\deltab)}}\big). 
\label{Eq:PartialTimesInverseCovariance}
\end{align}
Taking the quadratic form of this matrix with a vector of the perturbations \mbox{$(\deltapi,\deltab)$} gives the squared Mahalanobis distance (Equation~\ref{eq:MahaMom}).  

Now we want a matrix \mbox{$\mat{S}_{(\deltapi,\deltab) \leftarrow \boldsymbol{\eta}}^{ }$} that transforms a vector $\boldsymbol{\eta}$ of i.i.d.~normal entries into a random vector of perturbations. 
For $\deltab$, we only need to put the square root of the variances in Equation~\ref{Eq:DeltaB} along the corresponding diagonal of $\mat{S}$.  
For $\deltapi$, we need the individual realizations of the perturbations to sum to zero, so we put a projection matrix \mbox{$(\mat{I}-\frac{1}{k}\mathbf{1})/\sqrt{n k}$} along the corresponding diagonal of $\mat{S}$.\footnote{For the case of uniform \mbox{$\boldsymbol{\pi}_0^{ } = \frac{1}{k}$}, multiplying this projection matrix by $k$ i.i.d.~normal entries gives a vector of $k$ values that sum to zero, with covariance given by Equation~\ref{Eq:DeltaPi}.}

To obtain the desired matrix for moments $\mat{M}_{\vphantom{1}}^{\ExpLeftPar\mu\ExpRightPar}$, we multiply the matrix from Equation~\ref{Eq:PartialTimesInverseCovariance} on 
both sides by $\mat{S}$: 
\begin{align}
    \mat{M}_{\vphantom{1}}^{\ExpLeftPar\mu\ExpRightPar} = \big(\mat{S}_{(\deltapi,\deltab) \leftarrow \boldsymbol{\eta}}^{ }\big)^\top_{\vphantom{1}} \big(\mat{J}_{\partial{\mu}/\partial {(\deltapi,\deltab)}}\big)^\top_{\vphantom{1}}
\big(\mat{\Sigma}_{\vphantom{1}}^{\ExpLeftPar\mu\ExpRightPar}\big)^{-1}_{\vphantom{1}} \big(\mat{J}_{\partial{\mu}/\partial {(\deltapi,\deltab)}}\big) \big(\mat{S}_{(\deltapi,\deltab) \leftarrow \boldsymbol{\eta}}^{ }\big). 
\end{align}
The matrix for cumulants $\mat{M}_{\vphantom{1}}^{\ExpLeftPar\kappa\ExpRightPar}$ is defined analogously.  

As the $\pare$ is a ratio, it is natural to take the log before taking the expectation over the random perturbations. 
In particular, we can write the \mbox{$\langle \log \pare \rangle$} as follows: 
\begin{align*}
    \Big\langle \log \pare \Big\rangle &= \Big\langle \log \pae_{\vphantom{1}}^{\ExpLeftPar\kappa\ExpRightPar} \Big\rangle - \Big\langle \log \pae_{\vphantom{1}}^{\ExpLeftPar\mu)} \Big\rangle \\
    &= \Big\langle \log \big( \boldsymbol{\eta}^\top_{\vphantom{1}} \mat{M}_{\vphantom{1}}^{\ExpLeftPar\kappa\ExpRightPar} \boldsymbol{\eta} \big) \Big\rangle - \Big\langle \log \big( \boldsymbol{\eta}^\top_{\vphantom{1}} \mat{M}_{\vphantom{1}}^{\ExpLeftPar\mu\ExpRightPar} \boldsymbol{\eta} \big) \Big\rangle,
\end{align*}
where expectation is taken over $\boldsymbol{\eta}$ having i.i.d.~normal entries.   

\begin{figure}[H]
\begin{center}
\includegraphics[width=1.01\columnwidth]{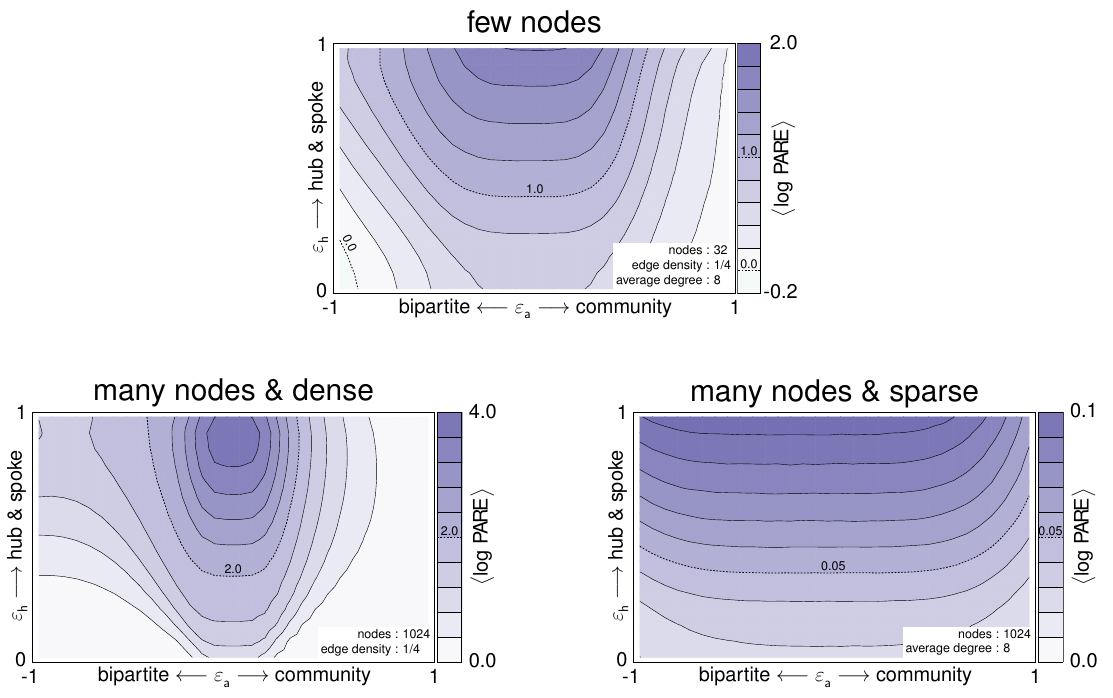}
\vspace{-2pt}
\caption{
\mbox{\textbf{The test using graph cumulants outperforms that using moments,}} \hspace{\textwidth} 
\mbox{\textbf{even in the limit of many observed graphs.}} \mbox{\hspace{\textwidth} \hspace{\textwidth}}
\mbox{Here,} we compare the asymptotic efficiencies of the two tests (with $r=3$) in the limit \mbox{$s\rightarrow\infty$}.  
The contours correspond to the \mbox{$\langle \log \pare \rangle$}
for random perturbations to the SBM with \mbox{$k=4$} equal-sized communities (Equation~\ref{eq:BforPARE}).
The test using cumulants does better than that using moments in nearly all regimes (aside from a small region for $32$ nodes).
}  
\label{fig:LogPARE} 
\end{center}
\end{figure}

As the distribution for $\boldsymbol{\eta}$ is rotationally symmetric, the entries are i.i.d.~normal for any orthogonal basis.
In particular, we diagonalize the (positive semidefinite) matrices $\mat{M}_{\vphantom{1}}^{\ExpLeftPar\mu\ExpRightPar}$ and $\mat{M}_{\vphantom{1}}^{\ExpLeftPar\kappa\ExpRightPar}$. 
Thus, we only need their eigenvalues $\lambda_i^{ }$ to compute the expected log $\pare$:
\begin{align}
    \Big\langle \log \pare \Big\rangle &= \Big\langle \log \sum_i^{ } \lambda_i^{\ExpLeftPar\kappa\ExpRightPar} \eta_i^2 \Big\rangle - \Big\langle \log \sum_i^{ } \lambda_i^{\ExpLeftPar\mu\ExpRightPar} \eta_i^2 \Big\rangle. \label{Eq:LogPAREFinal}
\end{align}
At this point, we estimate \mbox{$\langle \log \pare \rangle$} via Monte Carlo sampling, resulting in Figure~\ref{fig:LogPARE}.

\subsection{The Results}
\label{ssec:pareresult}

\vspace{2pt}

\newpage
\section{\mbox{The Limit of Few Observed Graphs} \mbox{ (Singular Covariance Estimates)}}
\label{Appendix:CovS}
Below a certain number of observed graphs, the test using moments results in singular covariance matrices.  
If the sum of the two covariance matrices is also singular, one cannot take its inverse, so the Mahalanobis distance (Equation~\ref{eq:MahaMom}) is \mbox{ill-defined}. 
Here, we illustrate the simplest example: when there is only \mbox{$s=1$} graph per sample, 
the distributions inferred by the test using moments have zero variance in the edge density $\mu_{\oneedge}^{ }$.  

Consider a single observed graph $G_{1}^{ }$ with $n$ nodes and $m$ edges.  
For the test using moments, the expected subgraph densities of the inferred distribution are taken to be those of this observed graph: \mbox{$\mu_g^{ }(\mathcal{G}) = \mu_g^{ }(G_{1}^{ })$}.  
The estimation of the variance of $\mu_{\oneedge}^{ }$ assumes that the graph moments match to second order (and, as $n$ is fixed, their expected counts as well).  
At first order, we have 
\begin{align}
    \langle c_{\oneedge}^{ }(G)\rangle_{G\sim\mathcal{G}}^{ } &= c_{\oneedge}^{ }(G_{1}^{ }) \nonumber\\
    &= 2m.  \label{eq:edgeexample}
\end{align}
And at second order, we have 
\begin{align}
    \langle c_{\oneedge}^{2}(G)\rangle_{G\sim\mathcal{G}}^{ } &= \langle 2 c_{\oneedge}^{ }(G) + 4c_{\twowedge}^{ }(G) + c_{\twoparallel}^{ }(G)\rangle_{G\sim\mathcal{G}}^{ } \nonumber\\
    &= 2c_{\oneedge}^{ }(G_{1}^{ }) + 4c_{\twowedge}^{ }(G_{1}^{ }) + c_{\twoparallel}^{ }(G_{1}^{ }) \nonumber\\
    &= \smash{(2m)^2_{\vphantom{1}}}.\label{eq:covedgexample}
\end{align}
The \textit{only} distributions satisfying both of these constraints are those containing graphs with precisely $m$ edges,\footnote{In Equations~\ref{eq:edgeexample} and~\ref{eq:covedgexample}, we have $2m$ edges instead of $m$ because we are considering (injective) homomorphism counts, which count both orientations of each edge.} and thus the variance in edge density is zero: \mbox{$\textrm{Var}(\mu_{\oneedge}^{ })=0$}.

In a sense, this can be thought of as a \mbox{$1$$\times$$1$} covariance matrix of rank $0$.  
In general, as the number of graphs per sample $s$ increases, the rank of the covariance matrix $\boldsymbol{\Sigma}^{\ExpLeftPar\mu\ExpRightPar}_{\vphantom{1}}$ does as well.  
For a given order $r$, one has covariance matrices of size \mbox{$|\boldsymbol{g}|$$\times$$|\boldsymbol{g}|$} (where $|\boldsymbol{g}|$ is the number of connected subgraphs with $r$ edges), and therefore requires sufficiently many graphs per sample $s$ such that the rank of the sum of the two inferred covariance matrices is no less than $|\boldsymbol{g}|$.
This is the reason behind the \mbox{$s<4$} cutoff for the test using moments in Figure~\ref{Fig:RealDataComparisonS}.  

We remark that, even with sufficient sample size, the covariance matrix may be singular (if a subset of the sample exhibits coincidental collinearity or any of its \mbox{higher-dimensional} analogues). 
To handle these infrequent cases in a consistent way, we use the \textit{pseudo}inverse in Equations~\ref{eq:MahaMom} and~\ref{eq:MahaCum}.

\newpage
\section{\mbox{Zero covariance between the edge density and} \mbox{ the other graph cumulants for \mbox{Erd\H{o}s-R\'enyi} distributions.}}
\label{Sec:ERCumCov}

\begin{theorem}
\label{thm:zerocovEdgeAndOthersForCumulantsER} 
For any \mbox{Erd\H{o}s-R\'enyi} distribution \mbox{$\ER(n, p)$}, the covariance between the edge density and all other unbiased estimators of graph cumulants \mbox{$\widehat{\kappa}_{g}^{ }$} is zero.
\end{theorem}

\begin{proof}
\label{proofzerocovEdgeAndOthersForCumulantsER}
We start by obtaining an expression for the covariance $\text{Cov}^{\ExpLeftPar\ER\ExpRightPar}_{\vphantom{1}}\big(\widehat{\mu}_{\oneedge}^{}(G), \widehat{\mu}_{g}^{}(G)\big)$ between the edge density\footnote{Recall that the edge density is both the first graph moment and the first graph cumulant, i.e.,\\ \mbox{$\mu_{\oneedge}^{ }(\mathcal{G})=\kappa_{\oneedge}^{ }(\mathcal{G})=\big\langle \widehat{\mu}_{\oneedge}^{ }(\boldsymbol{G}) \big\rangle_{\boldsymbol{G}\sim\mathcal{G}}^{ } = \big\langle \widehat{\kappa}_{\oneedge}^{ }(\boldsymbol{G}) \big\rangle_{\boldsymbol{G}\sim\mathcal{G}}^{ }$}.} and the moment of any subgraph $g$,
for graphs $G$ sampled 
from an \mbox{Erd\H{o}s-R\'enyi} distribution \mbox{$\ER(n, p)$}, where $n$ is the number of nodes in  $G$ and $p$ is the i.i.d.~probability of an edge between any pair of nodes.

Let $v$ denote the number of nodes in the subgraph $g$, and $e$ its number of edges. 
Recall that to compute the covariance between two subgraphs (Equation~\ref{Eq:Covariance}), one must consider all the ways that the nodes of the subgraphs could overlap (Section~\ref{SubSec:StatTestCov}). 
Here, there are four cases to consider: \\[-17pt]
\begin{enumerate}
    \item Neither node in the edge maps to any node in $g$. \\
    There is a single way this can happen, and it results in\\
    a subgraph with \mbox{$v+2$} nodes and \mbox{$e+1$} edges.
    \\[-17pt]
    \item A single node in the edge maps to a single node in $g$.\\ 
    There are $2v$ ways this can happen, each 
    resulting in\\a subgraph with \mbox{$v+1$} nodes and \mbox{$e+1$} edges.
    \\[-17pt]
    \item The two nodes of the edge map to two nodes in $g$ that do not already have an edge.  
    There are \mbox{\smash{$2\big[ {\tbinom{v}{2}v} - e\big] = v(v-1) - 2e$}} ways this can happen, each resulting in\\
    a subgraph with \mbox{$v$} nodes and \mbox{$e+1$} edges.
    \\[-17pt]
    \item The two nodes of the edge map to two nodes in $g$ that already have an edge.\\
    There are $2e$ ways that this can happen, each resulting in\\
    a subgraph with \mbox{$v$} nodes and \mbox{$e$} edges.
    \\[-17pt]
\end{enumerate}

For an \mbox{$\ER(n, p)$} distribution, the moment of a subgraph $g$ is $p^e$ (i.e., $p$ to the power of the number of edges in $g$). 
Thus, the moments of the resulting subgraphs in cases 1, 2, and 3 are \mbox{$p^{e+1}$}, while the moments of the resulting subgraphs in case 4 are \mbox{$p^{e}$}.  
Moreover, recall that the injective homomorphism counts \mbox{\smash{$c_{g}^{ }(G)$}} of a subgraph $g$ (with $v$ nodes) into a graph $G$ (with $n$ nodes) is \mbox{\smash{$\fallingfact{n}{v}\mu_{g}^{ }(G)$}}, where the falling factorial is 
\mbox{$\fallingfact{n}{v} = \prod_{j=0}^{v-1}(n-j)$.}

We now substitute these results into the expression for the covariance:
\begin{align*}
    \text{Cov}^{\ExpLeftPar\ER\ExpRightPar}_{\vphantom{1}}\big(\widehat{\mu}_{\oneedge}^{}(G), \widehat{\mu}_{g}^{}(G)\big)_{}  &= \big\langle\widehat{\mu}_{\oneedge}^{ }(G) \widehat{\mu}_{g}^{ }(G)\big\rangle_{G\sim\ER(n,p)}^{} - \big\langle\widehat{\mu}_{\oneedge}^{ }(G)\big\rangle_{G\sim\ER(n,p)}^{}\big\langle\widehat{\mu}_{g}^{ }(G)\big\rangle_{G\sim\ER(n,p)}^{} \nonumber\\
    &= \big\langle\widehat{\mu}_{\oneedge}^{ }(G) \widehat{\mu}_{g}^{ }(G)\big\rangle_{G\sim\ER(n,p)}^{} - \big(p\big)\big(p^{e}\big)\nonumber\\
    &= \frac{1}{\fallingfact{n}{v}\fallingfact{n}{2}}\bigg[ \big(1\big)\big(\fallingfact{n}{v+2}\big)\big(p^{e+1}\big) +\big(2v\big)\big(\fallingfact{n}{v+1}\big)\big(p^{e+1}\big)\\[-10pt]
    & \kern48pt +  \big(v(v-1) - 2e\big)\big(\fallingfact{n}{v}\big)\big(p^{e+1}\big)
    + \big(2e\big)\big(\fallingfact{n}{v}\big)\big(p^e\big)
    \bigg] - p^{e+1}\nonumber
\end{align*}
Using ``algebra autopilot'' and grouping the terms, we get: 
\begin{align}
    & 
     \frac{1}{\fallingfact{n}{2}}\bigg[ \Big( (n-v)(n-(v+1)) + 2v(n-v) + \big(v(v-1)-2e\big) - \fallingfact{n}{2}\Big)p^{e+1}  
     + (2e)p^{e}\bigg] \nonumber\\
     =&    \frac{1}{\fallingfact{n}{2}}\bigg[ \Big( n^2 - 2nv - n + v^2 + v + 2nv - 2v^2 + v^2- v -2e  - n^2 + n \Big)p^{e+1}  
     + (2e)p^{e}\bigg]  \nonumber\\
     =&    \frac{1}{\fallingfact{n}{2}}\bigg[ ( - 2e )p^{e+1}   + (2e)p^{e}\bigg]  
        =    \frac{1}{\fallingfact{n}{2}}    2e p^{e}\big(1 -  p\big)    \nonumber
\end{align}
\\[-17pt]
\begin{align}
\boxed{\vphantom{\binom{n}{2}}\kern6pt\text{Cov}^{\ExpLeftPar\ER\ExpRightPar}_{\vphantom{1}}\big(\widehat{\mu}_{\oneedge}^{ }, \widehat{\mu}_{g}^{ }\big)_{} \kern5pt = \kern6pt \underbrace{p\big(1-p\big)/\tbinom{n}{2}}_{\text{``variance of }\widehat{p}\!\text{ ''}} \kern7pt\times \underbrace{e p^{e-1}\vphantom{\tbinom{n}{2}}}_{\text{``}d(p^e)/dp\text{''}}} \label{Eq:CovEdgeERothersMu}
\end{align}

Crucially, for an $\ER(n,p)$ distribution, the covariance between the edge density and the density of a subgraph $g$ depends \textit{only} on the number of edges in $g$.

The \mbox{$\text{Cov}^{\ExpLeftPar\ER\ExpRightPar}_{\vphantom{1}}\big(\widehat{\kappa}_{\oneedge}^{}(G), \widehat{\kappa}_{g}^{}(G)\big)_{}\equiv\text{Cov}^{\ExpLeftPar\ER\ExpRightPar}_{\vphantom{1}}\big(\widehat{\mu}_{\oneedge}^{}, \widehat{\kappa}_{g}^{}\big)_{}$} 
is given by the sum of the covariances of the edge density with each subgraph $g'$ that appears in the expression for the unbiased 
cumulant of $g$ (weighted by its coefficient in the formula). 
For example:
\begin{align}
\text{Cov}^{\ExpLeftPar\ER\ExpRightPar}_{\vphantom{1}}\big(\widehat{\mu}_{\oneedge}^{}, \widehat{\kappa}_{\twowedge}^{}\big)_{} &= \text{Cov}^{\ExpLeftPar\ER\ExpRightPar}_{\vphantom{1}}\big(\widehat{\mu}_{\oneedge}^{}, \widehat{\mu}_{\twowedge}^{}\big)_{}  -
\text{Cov}^{\ExpLeftPar\ER\ExpRightPar}_{\vphantom{1}}\big(\widehat{\mu}_{\oneedge}^{}, \widehat{\mu}_{\twoparallel}^{}\big)_{}; \nonumber\\
\text{Cov}^{\ExpLeftPar\ER\ExpRightPar}_{\vphantom{1}}\big(\widehat{\mu}_{\oneedge}^{}, \widehat{\kappa}_{\threetriangle}^{}\big)_{} &= \text{Cov}^{\ExpLeftPar\ER\ExpRightPar}_{\vphantom{1}}\big(\widehat{\mu}_{\oneedge}^{}, \widehat{\mu}_{\threetriangle}^{}\big)_{} -
 3\kern1.5pt\text{Cov}^{\ExpLeftPar\ER\ExpRightPar}_{\vphantom{1}}\big(\widehat{\mu}_{\oneedge}^{}, \widehat{\mu}_{\threeedgewedge}^{}\big)_{} + 
 2\kern1.5pt\text{Cov}^{\ExpLeftPar\ER\ExpRightPar}_{\vphantom{1}}\big(\widehat{\mu}_{\oneedge}^{}, \widehat{\mu}_{\threeparallel}^{}\big)_{}; 
 \nonumber \\
\text{Cov}^{\ExpLeftPar\ER\ExpRightPar}_{\vphantom{1}}\big(\widehat{\mu}_{\oneedge}^{}, \widehat{\kappa}_{\threeline}^{}\big)_{}  &= 
\text{Cov}^{\ExpLeftPar\ER\ExpRightPar}_{\vphantom{1}}\big(\widehat{\mu}_{\oneedge}^{}, \widehat{\mu}_{\threeline}^{}\big)_{}  -
 2\kern1.5pt\text{Cov}^{\ExpLeftPar\ER\ExpRightPar}_{\vphantom{1}}\big(\widehat{\mu}_{\oneedge}^{}, \widehat{\mu}_{\threeedgewedge}^{}\big)_{}  + 
\text{Cov}^{\ExpLeftPar\ER\ExpRightPar}_{\vphantom{1}}\big(\widehat{\mu}_{\oneedge}^{}, \widehat{\mu}_{\threeparallel}^{}\big)_{}. \nonumber 
\end{align}
We now note two key facts about the expressions for the unbiased graph cumulants: \textit{1)} the subgraphs all have the same number of edges, and \textit{2)} the coefficients always sum to zero.
Hence, \mbox{$\text{Cov}^{\ExpLeftPar\ER\ExpRightPar}_{\vphantom{1}}\big(\widehat{\mu}_{\oneedge}^{}, \widehat{\kappa}_{g}^{}\big)_{}=0$} for any other subgraph $g$, 
as desired. 
\end{proof}

\newpage
\vskip 0.2in
\bibliography{referencesgraphcumulants}

\end{document}